\documentclass[
reprint,
amsmath,amssymb,
aps,physrev,
floatfix
]{revtex4-2}

\usepackage{graphicx}   
\usepackage{dcolumn}    
\usepackage{bm}         
\usepackage{hyperref}   
\usepackage[export]{adjustbox}

\begin{document}

\preprint{APS/123-QED}

\title{\textbf{Universal BPS Structure of Scalar Kinks in Static Geometries} 
}%

\author{G. Luchini}
 \email{gabriel.luchini@ufes.br}
\author{G. B. Sant'Anna}%
 \email{georgia.battisti@ufes.br}
\author{U. Camara da Silva}%
 \email{ulysses.silva@ufes.br}
\affiliation{
UniversidSe Federal do Esp\'irito Santo – Departamento de F\'isica\\
Av. Fernando Ferrari, Goiabeiras, 29075-900, Vit\'oria-ES, Brasil}



\date{\today}

\begin{abstract}
We present a geometric extension of the Bogomolny–Prasad–Sommerfield (BPS) construction for scalar kinks in $(1+1)$ dimensions embedded in static curved spacetimes. By introducing a nonminimal coupling between the scalar prepotential and the extrinsic curvature of the static foliation, the flat-space first-order Bogomolny equation remains exactly valid for arbitrary static backgrounds. As a consequence, the kink profile is unchanged, while the effective potential and vacuum structure acquire a controlled geometric dependence. We show that these curved-space BPS kinks are always linearly stable. However, the existence of the translational zero mode is not guaranteed: its normalizability depends on the competition between the intrinsic length scale of the kink and the asymptotic curvature scale of the geometry. When the geometric scale dominates, the zero mode is removed and the soliton becomes geometrically pinned, despite remaining an exact BPS solution. Explicit realizations in $\mathrm{AdS}_2$ demonstrate how different static slicings of the same spacetime lead to qualitatively distinct physical outcomes, ranging from preserved translational invariance to its complete removal by horizons. These results establish geometry as a precise mechanism for controlling solitonic moduli without compromising linear stability.

\end{abstract}

\maketitle


\emph{Introduction --}  Topological solitons — kinks and antikinks — in $(1+1)$-dimensional relativistic field theories in flat space-time provide some of the simplest settings in which nonperturbative phenomena can be analyzed exactly, offering insights into stability, duality, and quantization \cite{Rajaraman,MantonSutcliffe,Bazeia,Agostinho}. In a broad class of models for a real scalar field $\phi(t,x)$, the dynamics is governed by a potential $\widetilde V(\phi)$ possessing a discrete set of degenerate vacua $\{\phi_n\}$, $n=1,2,\dots$, for which $\widetilde V(\phi_n)=0$.  When the potential can be expressed in terms of a prepotential $\mathcal Q(\phi)$ as
\begin{equation}
\widetilde V(\phi)=\tfrac{1}{2}\big(\partial_\phi\mathcal Q(\phi)\big)^2,
\end{equation}
the model admits minimum-energy kink solutions that satisfy the first-order Bogomolny–Prasad–Sommerfield (BPS) equation\cite{Bogomolny,PrasadSommerfield}
\begin{equation}
\label{BPS}
\partial_x\phi(x)=\pm\,\partial_\phi\mathcal Q(\phi).
\end{equation}
These BPS configurations saturate a topological energy bound, automatically solve the full second-order equations of motion, and are linearly stable.  The
latter property is reflected in their fluctuation spectrum, which is governed by a supersymmetric quantum-mechanical (SUSY-QM) structure~\cite{Cooper}.

First-order BPS-like equations do appear in gravitational contexts, notably in holographic \cite{Maldacena,witten1998antisitterspaceholography} studies of renormalization group flows\cite{Freed_Gub} and within the fake supergravity formalism \cite{PRL_Sken_Town,fake_super}. However, when gravity is dynamical, these are not genuine BPS bounds: the corresponding solutions do not minimize an action, and the vacuum structure is constrained by Einstein's equations, obscuring the intrinsic scalar dynamics. 

This raises a fundamental question: does the BPS structure of flat-space kinks persist universally in fixed curved background, or is it inherently tied to flatness?

We give an answer to this question by presenting a universal mechanism that extends any flat-space BPS model to an arbitrary static curved geometry, preserving the exact same first-order Bogomolny equation. The key is a specific non-minimal coupling of the scalar field to the extrinsic curvature of the space-time foliation. This coupling geometrically deforms the potential while leaving the kink's spatial profile untouched. The construction includes an essential boundary term that keeps the action finite and drives the BPS energy to zero, so that the vacuum-interpolating character of the kink is encoded in the solution’s boundary conditions rather than in its energy.

This geometric embedding has important consequences. While the kink remains an ordinary BPS solution, the curvature can dramatically change its physical properties. Most notably, we derive a simple geometric criterion that determines whether the translational zero mode survives. When the asymptotic curvature exceeds the vacuum mass scale, the zero mode becomes non-normalizable. The kink is effectively pinned by the geometry, and supersymmetry in the fluctuation spectrum is spontaneously broken. We thus establish a robust framework for BPS solitons in curved space, revealing a new interplay between topology and geometry that controls the fate of moduli and stability.

\emph{Preserving the BPS Equation in Curved Space --} The standard Bogomolny construction implicitly relies on the presence of a global inertial frame in which the kink can be regarded as static. In curved or accelerated settings no such frame exists, making it nontrivial to preserve a first–order description. We now present a universal prescription for embedding any flat-space BPS kink into a static curved spacetime of the form
\begin{equation}
    ds^{2}=h^{2}(x)\, dt^{2}-dx^{2},
    \label{metric}
\end{equation}
where the lapse function $h(x)$ encodes intrinsic geometric properties of the background and defines the corresponding static foliation.  Even on the same underlying manifold, different choices of $h(x)$ highlight different geometric aspects of the spacetime and represent inequivalent physical observers.

\begin{figure}[t]
    \centering
    \includegraphics[width=0.8\columnwidth]{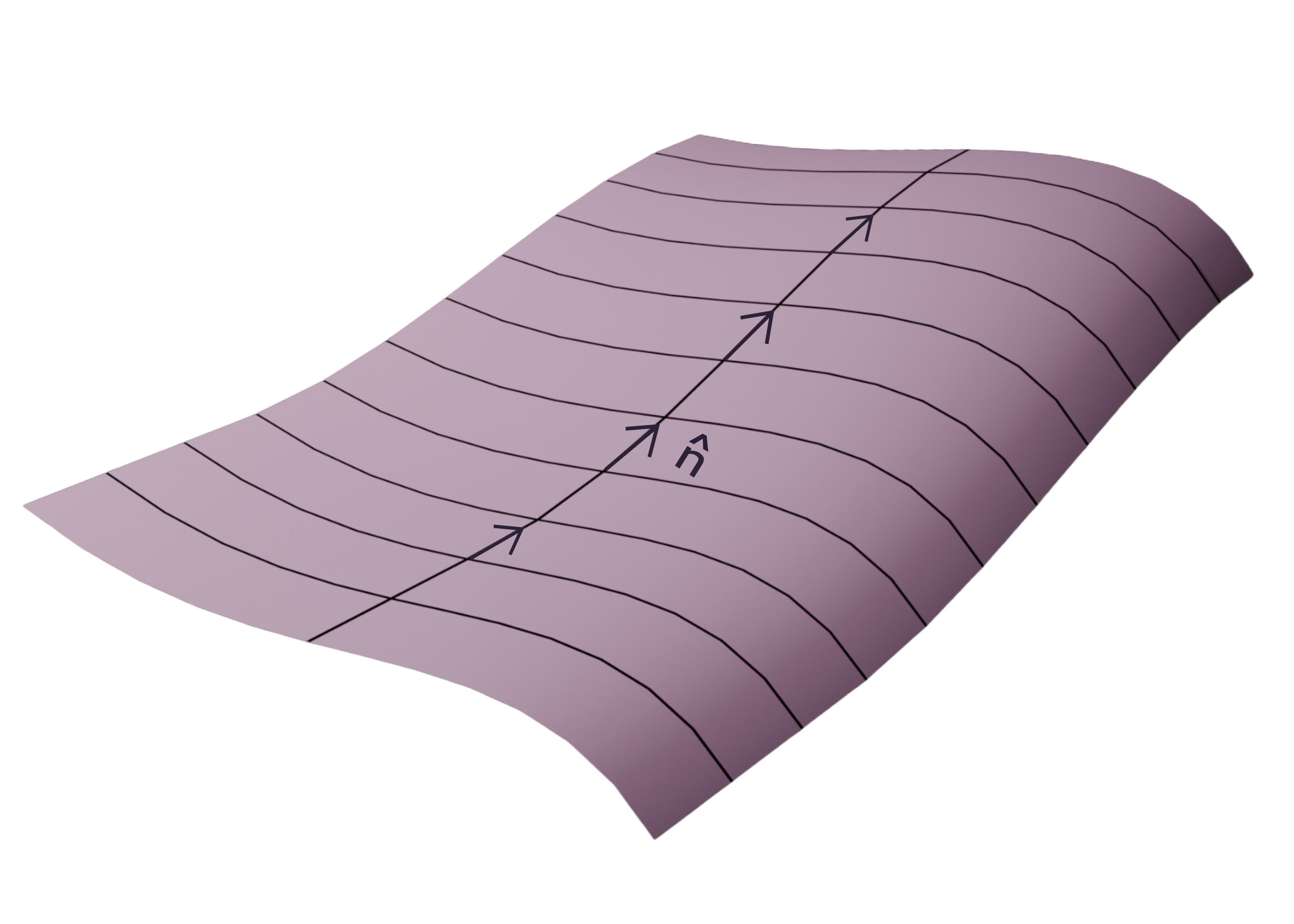}
    \caption{
Geometric interpretation of a static BPS configuration in curved spacetime. A preferred spatial path, defined by the unit normal vector of a chosen static foliation, is selected so that the soliton can remain static despite curvature.
}

    \label{fig1}
\end{figure}

When a scalar field is embedded in a curved background of this form, the notion of a static configuration requires specifying, at each point, the spatial direction along which the field profile is traced.  Because the relation between temporal and spatial directions changes locally through the function $h(x)$, the kink cannot remain static unless its profile is aligned with the spacelike normal $n^\mu=(0,1)$ of the chosen foliation, which defines the spatial direction along which the field varies, as depicted in Fig.\ref{fig1}. For the first-order Bogomolny equation to survive this geometric distortion, the dynamics must incorporate a compensating term that tracks exactly how the preferred direction bends. The variation of the unit normal vector along the spacetime is captured by the extrinsic curvature
\begin{equation}
K(x)=\nabla_\mu n^\mu=\partial_x\ln h(x),
\end{equation}
which measures how the worldlines $x=\mathrm{const}$ deviate from spacetime geodesics \cite{WaldGR}. Equivalently, $K$ coincides with the proper acceleration required to keep an observer at fixed $x$: when $K=0$ the observer is in free fall, while $K\neq0$ signals non-geodesic, accelerated motion. The nonminimal coupling to $K(x)$ accounts for the curvature of the spacelike normal direction, ensuring that the flat-space Bogomolny equation remains valid along the kink profile.

A crucial feature of the BPS structure is that the shift transformation of the pre-potential by a constant, $\mathcal{Q}\!\to\!\mathcal{Q}+c$, is a symmetry of the flat-space action. A curved-space generalization that aims to preserve the BPS equation \eqref{BPS} must therefore respect this symmetry.
The action
\begin{align}
S &= \int_{\Sigma} d^2x \,\sqrt{-g}\left[
       \frac{1}{2}g^{\mu\nu}\partial_\mu\phi\,\partial_\nu\phi 
       - V_{\pm}(\phi,x)\right]
   + S_{B}^{(\pm)}, 
   \label{action-curved}\\[2mm]
S_{B}^{(\pm)} &= 
   \mp \int_{(\partial\Sigma)_n} dx\,\sqrt{-\gamma}\;\mathcal{Q}(\phi),
\label{boundary-term}
\end{align}
where $\gamma$ is the induced metric on the boundary, implements 
this requirement with  a potential that is modified by a nonminimal geometric
coupling between the field and the extrinsic curvature $K(x)$,
\begin{equation}
    V_{\pm}(\phi,x)
    =\frac{1}{2}\big(\partial_\phi\mathcal{Q}(\phi)\big)^{2}
      \pm K(x)\,\mathcal{Q}(\phi),
    \label{V-curved}
\end{equation}
and this is a deformation that both preserves the shift symmetry, since under
$\mathcal{Q}\!\to\!\mathcal{Q}+c$, the bulk term transform as
\[
\sqrt{|g|}\,K(x)\mathcal{Q} \;\longrightarrow\;
\sqrt{|g|}\,K(x)\mathcal{Q} + c\,\partial_x h(x),
\]
acquiring a total derivative that is exactly canceled by the boundary contribution $S_{B}^{(\pm)}$, so rendering the action invariant, and leaves the first-order BPS equation unchanged, as can be directly seen once the action \eqref{action-curved} is written with the metric \eqref{metric} as

\begin{equation}
    S=\int d^2x\,h(x)\left[\frac{1}{2h^2(x)}(\partial_t\phi)^2
         -\frac{1}{2}\left(\partial_x\phi \mp \partial_\phi\mathcal{Q}\right)^2\right],
    \label{action2}
\end{equation}
giving the static Euler-Lagrange equation
\begin{equation}
    \partial_x^2 \phi + K(x)\,\partial_x\phi
    = \partial_\phi\mathcal{Q}\partial_\phi^2\mathcal{Q} \pm K(x)\,\partial_\phi\mathcal{Q},
    \label{EOM}
\end{equation}
for which the first-order BPS equation remains an exact solution. The kink profile is identical to its flat-space counterpart, but it now exists in a curved background determined by $h(x)$.

 The corresponding static energy functional,
\begin{equation}
E=\frac{1}{2}\int_{\Sigma_t} dx\,h(x)\left(\partial_x\phi \mp \partial_\phi\mathcal{Q}\right)^2,
\end{equation}
is minimized precisely by the BPS solutions. This stands in contrast to the familiar Minkowski case, where a BPS kink has finite nonzero energy proportional to its topological charge,
\(\Delta\mathcal{Q} = \mathcal{Q}(\phi_{n+1}) - \mathcal{Q}(\phi_n)\).

In the flat case $K=0$, the boundary term in the action \eqref{action-curved} cancels the topological contribution $\Delta\mathcal{Q}$ to the energy, driving the BPS bound to zero.
The vacuum--interpolating character of the kink remains encoded in the field profile and in the mapping between distinct vacua, but it no longer
contributes additively to the energy because the nonminimal curvature coupling forces the BPS sector to be defined relative to the chosen static foliation.  Thus, the vanishing of the BPS energy in curved space (and in its flat limit) is not a defect of the construction but rather a consequence of the geometric mechanism that preserves the first--order equation while keeping the action finite.

\begin{figure}[t]
    \centering
    \includegraphics[width=0.8\columnwidth]{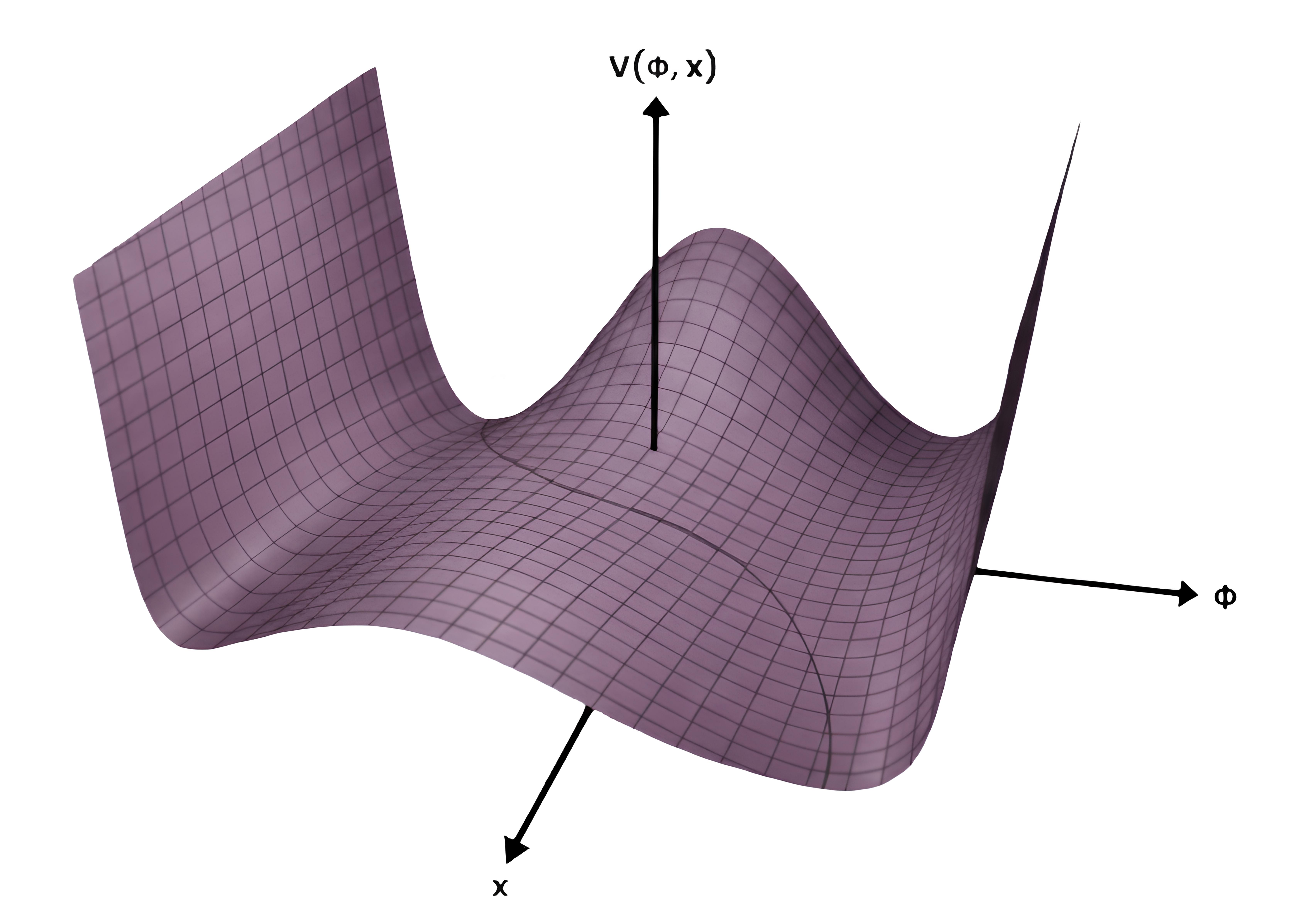}
   \caption{
Geometric deformation of the scalar potential induced by curvature. A nonminimal coupling between the scalar prepotential and the extrinsic curvature reshapes the effective potential in a position-dependent manner, while preserving the flat-space BPS equation.}

    \label{fig2}
\end{figure}

The structure of the vacua in curved space is correspondingly more subtle: given that
\begin{align}
\partial_\phi V_{\pm} &= \partial_\phi\mathcal{Q}
   \left(\partial_\phi^2\mathcal{Q} \pm K(x)\right),\\
\partial_\phi^2 V_{\pm} &= \partial_\phi^2\mathcal{Q}
   \left(\partial_\phi^2\mathcal{Q}\pm K(x)\right)
 + \partial_\phi\mathcal{Q}\,\partial_\phi^3\mathcal{Q},
\end{align}
the BPS vacua satisfying $\partial_\phi \mathcal{Q}(\phi_n)=0$ remain extrema of $V_\pm$ for all $x$, but their characterization as minima or maxima is controlled by the background geometry through

\begin{equation}
    M^2(x)\equiv 
    \partial_\phi^2\mathcal{Q}(\phi_n)
    \left(\partial_\phi^2\mathcal{Q}(\phi_n)\pm K(x)\right).
\end{equation}
Consequently, a kink solution need not interpolate between minima of $V_{\pm}$ at spatial infinity unless $M^2(\pm\infty)>0$. This scenario is represented in Fig.~\ref{fig2}.

Requiring the value of the extrema of the potential, and equivalently the effective vacuum mass $M^2(x)$, to remain finite in the asymptotic regions, forces the lapse function to exhibit exponential behavior at the ends of the spatial domain. Depending on the chosen static foliation, the asymptotic regions may correspond either to $\mathrm{AdS}$ boundaries or horizons, respectively characterized as
\begin{equation}
h(x) \propto e^{\kappa_b |x|}, \qquad h(x) \propto e^{-\kappa_h |x|},
\qquad \kappa_b , \kappa_h > 0,
\label{eq:asymptotic_h}
\end{equation}
The identification of a given asymptotic region as a boundary or a horizon is not fixed a priori by the sign of $x$, depending on the global behavior of the lapse function $h(x)$.

It is important to emphasize that the choice of the function $h(x)$ appearing in the static metric does more than specify coordinates: it determines the geometric direction along which the kink is static.  Each choice of static slicing therefore selects a preferred spatial direction, and the BPS equation holds precisely because it is formulated along this distinguished direction, which explicitly breaks full diffeomorphism invariance.  Moreover, the lapse function also fixes the domain of the spatial coordinate itself, since zeros $x_s$ near which 
\begin{equation}
\label{h}
h(x) \propto (x-x_s)^\alpha \left(1+ \sum_{n=1}a_n (x-x_s)^{n}\right), \,\alpha \geq 0, 
\end{equation}
may reflect on the existence of geometrical horizons ($0 \leq \alpha \leq 1$), or singularities ($\alpha > 1$).

\emph{Linear Stability and the Zero Mode --}
Having established that the BPS equation retains its flat-space form in a broad class of static curved geometries, it is natural to ask whether this robustness extends beyond the classical configuration itself.  In particular, the dynamical stability of the kink depends on the spectrum of small fluctuations around the BPS background, and geometric effects may alter the existence of the translational zero mode or even break the supersymmetric structure typically associated with BPS states. The most intriguing aspect of this geometric sensitivity is that the kink profile remains an exact BPS solution for all these geometries, yet the
existence of the zero mode is not guaranteed.  We now turn to the fluctuation analysis to show explicitly how curvature controls the behavior of the zero-energy sector.  We begin by considering the perturbed configuration

\begin{equation}
\phi(x,t)=\phi_{k,\bar{k}}(x)+\Psi(x,t),
\end{equation}
where $\phi_{k,\bar{k}}(x)$ correspond to kink ($k$) and anti-kink ($\bar{k}$) configurations which solve the BPS equation~(\ref{BPS}). Expanding the action to quadratic order in $\Psi$ gives
\begin{eqnarray}
\!\!\!\!\!\!\!\!\!S=\int d^2x 
\left[
\frac{1}{2h}(\partial_t\Psi)^2
-\frac{h}{2}\left(
\partial_x\Psi 
\mp 
\partial_\phi^2\mathcal{Q}\big|_{\phi_{k,\bar{k}}}\Psi
\right)^2
\right].
\label{S_flut}
\end{eqnarray}
Varying this action and introducing the coordinate 
\begin{equation}
z=\int^x \frac{dx'}{h(x')},
\end{equation}
the fluctuation equation becomes
\begin{eqnarray}
-\partial_z^2\Psi + \mathcal{V}(z)\Psi = -\partial_t^2\Psi, \\
\mathcal{V}(z)=
\left(h\,\partial_\phi^2\mathcal{Q}\big|_{\phi_{k,\bar{k}}}\right)^2
\pm 
\partial_z\left(h\,\partial_\phi^2\mathcal{Q}\big|_{\phi_{k,\bar{k}}}\right),
\label{flut_1}
\end{eqnarray}
with the upper (lower) signs referring to the kink (anti-kink).

Decomposing the fluctuations into Fourier modes, $\Psi(z,t)=e^{-i\omega t}\Psi_\omega(z)$, Eq.~(\ref{flut_1}) takes the supersymmetric quantum-mechanics form
\begin{align}
\mathcal{A}^\dagger\mathcal{A}\,\Psi_\omega = \omega^2 \Psi_\omega,
\label{flut_2}
\end{align}
where
\begin{align}
\mathcal{A} &= \partial_z \mp h(z)\,\partial_\phi^2\mathcal{Q}\big|_{\phi_{k,\bar{k}}}, \\
\mathcal{A}^\dagger &= -\partial_z \mp h(z)\,\partial_\phi^2\mathcal{Q}\big|_{\phi_{k,\bar{k}}}.
\end{align}
Since $\mathcal{A}^\dagger\mathcal{A}$ is a positive semi-definite operator, the condition for linear stability,
\begin{equation}
\omega^2 \ge 0,
\end{equation}
is automatically satisfied. Thus kink and anti-kink configurations remain linearly stable in any static curved background compatible with our BPS construction.

The normalizability of linear fluctuations
\begin{equation}
\label{zero_mode}
\int dz\,|\Psi_\omega(z)|^{2}
   =\int dx\,h^{-1}(x)\,|\Psi_\omega(x)|^{2}<\infty
\end{equation}
is controlled by the asymptotic behavior of $h^{-1}(x)\,|\Psi_\omega(x)|^{2}$, and the limits of integration are therefore defined through $h(x)$: zeros or singularities of the lapse can terminate the spatial interval, which need not coincide with $x \in (-\infty,+\infty)$, restricting the admissible range of~$x$ and potentially amputating the BPS configuration. 

The zero mode, when it exists, solves $\mathcal{A}\Psi_0=0$, giving
\begin{equation}
\Psi_0(z)\propto \partial_\phi\mathcal{Q}\big(\phi_k(z)\big),
\end{equation}
exactly as in flat space where this function is always normalizable, reflecting the exact translational symmetry $x\to x+a$ of the background. In curved space, however, such translations are generally not isometries, and normalizability of the zero mode is no longer guaranteed. What remains essential is the time-independence of the BPS equation, which ensures that a candidate zero mode exists, but not necessarily that it belongs to the physical Hilbert space. In what follows we analyze the conditions for their normalizability.

When the BPS profile is fully contained within the chosen coordinate patch, so that the geometry does not truncate the kink and the field approaches a vacuum $\phi(x)\approx\phi_n$ as $x\to\pm\infty$, with $h(x)\sim e^{\kappa_\pm x}$, the expansion of the prepotential leads to

\begin{equation}
\mathcal{Q}(\phi)\simeq \mathcal{Q}(\phi_n)
+ \frac{1}{2}\partial_\phi^2\mathcal{Q}\big|_{\phi_n}\,(\phi-\phi_n)^2,
\end{equation}
and the BPS equation \eqref{BPS} becomes linear,
\begin{equation}
\partial_x\phi = 
\pm\partial_\phi^2\mathcal{Q}\big|_{\phi_n}\,(\phi-\phi_n),
\end{equation}
with solution
\begin{equation}
\phi(x)-\phi_n\propto
\exp\!\left[
\pm\,\partial_\phi^2\mathcal{Q}\big|_{\phi_n}\,x
\right]\to0,
\end{equation}
which fixes the asymptotic sign of $\partial_\phi^2\mathcal{Q}$ at the connected vacua: it is positive at the $x\to-\infty$ end for a kink (and negative for an antikink), and the sign flips at $x\to+\infty$.

 The asymptotic behaviour of the integrand in \eqref{zero_mode} for $x\to \infty$, the $\mathrm{AdS}$ boundary, is given by 
\begin{equation}
h^{-1}(x)|\Psi_0|^2 \sim 
\exp\!\left[-\left(2\big|\partial_\phi^2\mathcal{Q}(\phi_{n+1})\big| +\kappa_+\right)x\right]
\end{equation}
and therefore, the normalizability is always fulfilled. For the $\mathrm{AdS}$ horizon, with $x\to -\infty$, one has
\begin{equation}
h^{-1}(x)|\Psi_0|^2 \sim 
\exp\!\left[-\left(2\big|\partial_\phi^2\mathcal{Q}(\phi_n)\big| -\kappa_-\right)|x|\right],
\end{equation}
and the normalizability requires
\begin{equation}
2\left|\partial_\phi^2\mathcal{Q}(\phi_n)\right| \equiv \frac{2}{\ell_n} > \kappa_-,
\label{normalizability}
\end{equation}
with $\ell_n$ determining the length scale of the BPS solution. 

When this inequality is violated, the curvature overwhelms the intrinsic mass scale ${\ell^{-1}_n}$ of the vacuum and the would-be translational mode ceases to be normalizable. 

For the cases where the BPS configuration is truncated by the range of the system of coordinates and we have geometrical horizons or singularities at the point $x_s$, the lapse function $h(x)$ behaves asymptotically as given in equation \eqref{h} and we get for $\alpha \neq 1$ and $\alpha = 1$ respectively that
\begin{equation}
\!\!\!\int^{x_s}h^{-1}(x)\vert \Psi_0\vert^2\;dx \sim
\left\{
\begin{array}{c}
\vert \Psi_0(x_s)\vert^2(x-x_s)^{1-\alpha} \\
\vert \Psi_0(x_s)\vert^2\ln{(x-x_s)}
\end{array}
\right.
\end{equation}
which shows that the zero mode is normalizable at $x_s$ when $0\leq \alpha <1$.

As a result, the continuous family of kink solutions parameterized by their position no longer exists: the translational modulus is removed, the kink becomes pinned to a fixed location determined by the geometry, and supersymmetry in the fluctuation spectrum is spontaneously broken despite the formal factorization of the Hamiltonian.

This curvature-induced removal of the translational degree of freedom highlights the central feature of our construction: a (anti-)kink, truncated by the system of coordinates or not, can remain BPS in curved space while losing its translational zero mode purely due to curvature effects. In the next section we demonstrate this mechanism explicitly in several coordinates systems of $\mathrm{AdS}_2$ background for the paradigmatic $\phi^4$-model.

\emph{BPS kinks of the $\phi^4$ model in $\mathrm{AdS}_2$ --} To illustrate the general geometric mechanism developed above, we now apply our construction to the standard $\phi^{4}$ model, whose prepotential is 
\begin{equation}
\mathcal{Q}(\phi)
= \frac{1}{2\ell\phi_0}\!\left(\phi_0^2\phi-\frac{1}{3}\phi^3\right),
\end{equation}
so that the BPS equation~\eqref{BPS} takes the form
\begin{equation}
\partial_x \phi
= \frac{1}{2\ell\phi_0}\left(\phi_0^2-\phi^2\right).
\end{equation}
The corresponding kink solution,
\begin{equation}
\label{kink}
\phi(x)=\phi_0\tanh\!\left(\frac{x-x_0}{2\ell}\right),
\end{equation}
provides the canonical example of a topological soliton in $(1+1)$-dimensional relativistic field theory. As discussed in the general analysis above, the zero mode for the kink solution, when it exists, is $\Psi_0(x)\propto \partial_x\phi_k(x)$. For the $\phi^4$-model we have explicitly
\begin{equation}
\Psi_0(x) \propto \frac{1}{\cosh^2\left[(x-x_0)/2\ell)\right]},
\label{phi4_zero}
\end{equation}
which coincides with the flat-space result. The physical relevance of this mode, however, depends on its normalizability with respect to the curved-space measure, a question that is controlled by the asymptotic behavior of the chosen $\mathrm{AdS}_2$ foliation.

Because of its universality and broad range of applications, from condensed matter systems \cite{Krumhansl1975} to cosmology \cite{Gabi_1}, the $\phi^4$ kink provides a well-controlled reference configuration for isolating the effects introduced by the background geometry.

As a curved background, we consider $\mathrm{AdS}_2$, defined as the hyperboloid
\begin{equation}
(X^0)^2-(X^1)^2+(X^2)^2=\frac{1}{\kappa^2},
\end{equation}
embedded in a three-dimensional space with line element
\begin{equation}
ds^2=(dX^0)^2-(dX^1)^2+(dX^2)^2.
\end{equation}
Besides Minkowski space, $\mathrm{AdS}_2$ is maximally symmetric and has constant negative curvature, making it a natural arena for testing the robustness of BPS solitons under geometric deformations. Moreover, $\mathrm{AdS}_2$ plays a central role in two-dimensional holography \cite{AdS_2_holo}, Jackiw–Teitelboim gravity \cite{Jackiw1985}, and the near-horizon physics of extremal black holes \cite{BardeenHorowitz}, providing additional motivation for this choice.

With appropriate parametrizations of the embedding coordinates $(X^0, X^1, X^2)$ on the $\mathrm{AdS}_2$ hyperboloid, the induced metric can be written in the static form~\eqref{metric} in several physically inequivalent ways. These different parametrizations correspond to distinct static foliations of the same underlying spacetime and lead to different lapse functions $h(x)$. In particular, $\mathrm{AdS}_2$ admits three natural static coordinate systems defined by
\begin{align}
h(x)&=\cosh(\kappa x), & K(x)&=\kappa\tanh(\kappa x), & x\in\mathbb{R}, \label{case1}\\
h(x)&=\sinh(\kappa x), & K(x)&=\kappa\coth(\kappa x), & x>0,\label{case2}\\
h(x)&=e^{\kappa x},    & K(x)&=\kappa,              & x\in\mathbb{R}. \label{case3}
\end{align}
As guaranteed by the construction, the BPS equation retains its flat-space form and the kink profile remains exactly as given in \eqref{kink}, while the potential is deformed by the curvature coupling. 

\begin{figure}[!h]
    \centering
    \includegraphics[width=0.48\columnwidth]{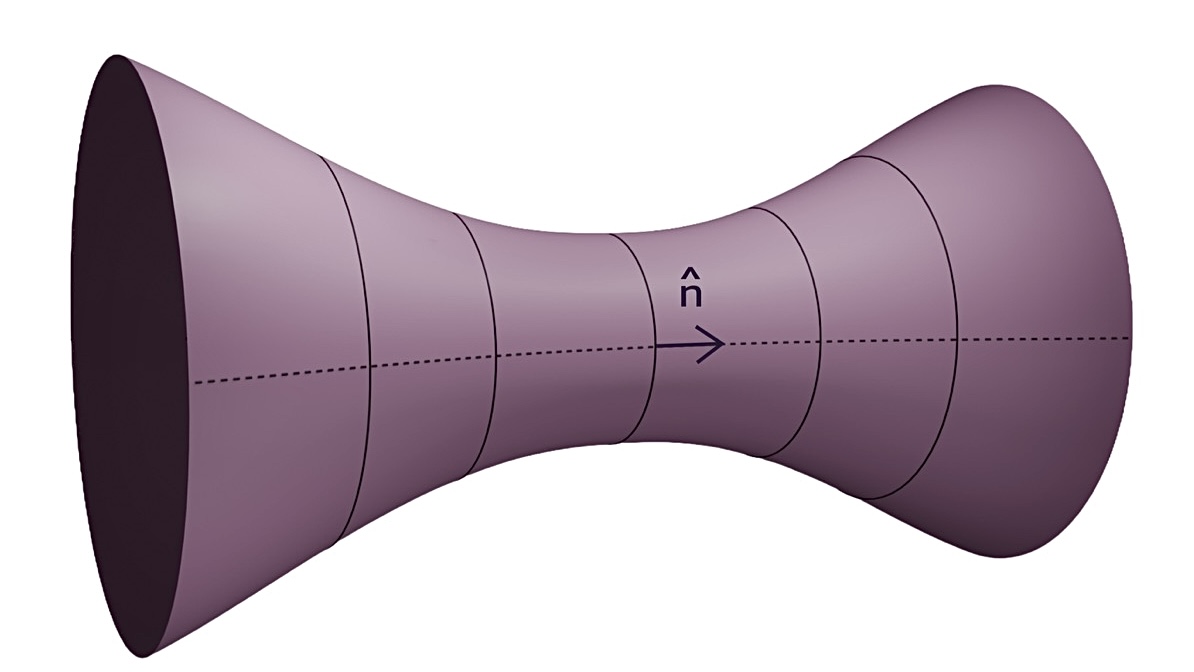}\hfill
    \includegraphics[width=0.48\columnwidth]{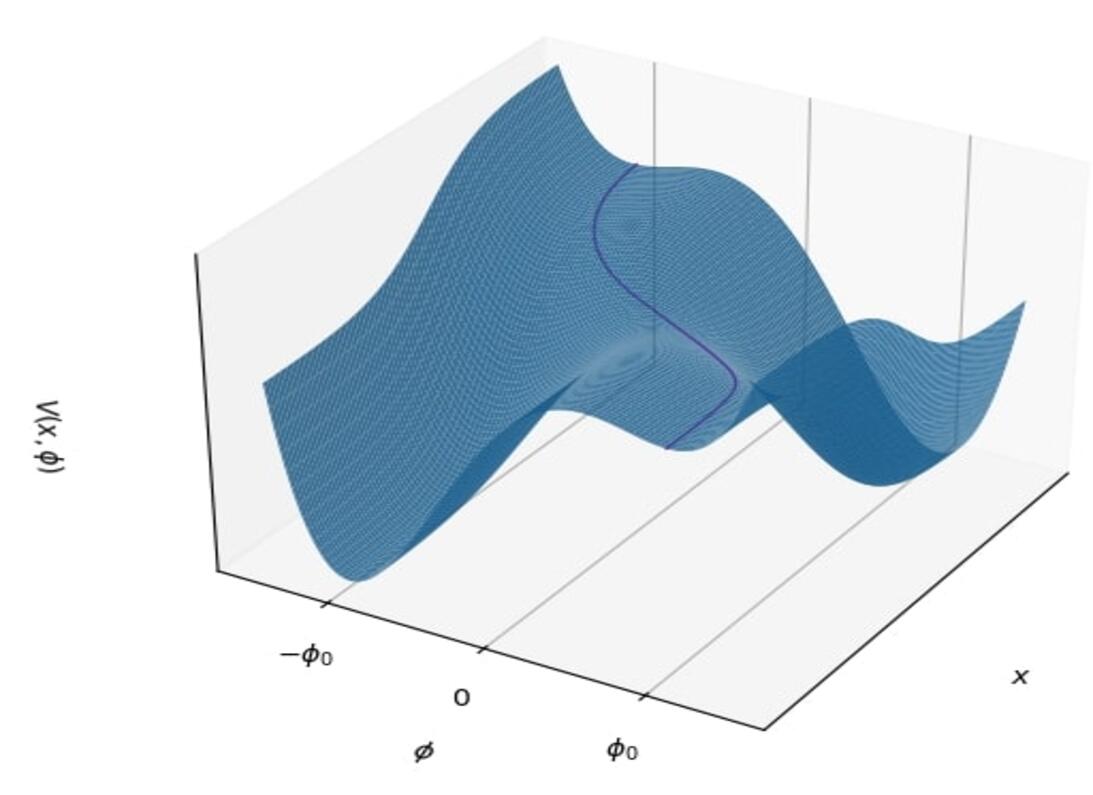}
    \caption{
Global static foliation of the $\mathrm{AdS}_2$ hyperboloid and BPS kink profile for $h(x)=\cosh(\kappa x)$. The translational zero mode remains normalizable despite the absence of translational isometries.
}

    \label{fig3}
\end{figure}

The slicing defined by the global coordinates, given in \eqref{case1}, covers the entire $\mathrm{AdS}_2$ space-time as depicted in Fig.\eqref{fig3}.

Linear fluctuations around the $\phi^4$-kink background are governed by the Schr\"odinger-type equation~\eqref{flut_1} and the fluctuation operator \eqref{flut_2} will depend explicitly on the parameters $\ell$ and $\kappa$.  In the global coordinates, the lapse function diverges exponentially as $|x|\to\infty$, so, as discussed in the general analysis, this asymptotic behavior guarantees that the translational zero mode is always square integrable and is given by equation \eqref{phi4_zero}.

A particularly interesting  case arises at the tuned value $\kappa=\frac{1}{2\ell}$, with $x_0 = 0$. Here, the coordinate transformation simplifies, giving the relation $\sinh{(\kappa x)} = \tan{(\kappa z)}$, and the fluctuation potential in \eqref{flut_1} reduces to
\begin{equation}
\mathcal{V}(z)
=-\frac{1}{\ell^2}+\frac{1}{2\ell^2}\sec^2(\kappa z),
\end{equation}
corresponding to the trigonometric P\"oschl–Teller potential on a finite interval $z \in \left(-\frac{\pi}{2\kappa}, \frac{\pi}{2\kappa}\right)$, \cite{PT_original,Uly_2}. The full discrete spectrum is therefore exactly solvable, with frequencies
\begin{equation}
\omega_n=\frac{1}{2\ell}\sqrt{n(n+4)},
\qquad n=0,1,2,\ldots,
\end{equation}
and normalized eigenfunctions
\begin{equation}
\Psi_n(z)=\mathcal{N}_n\,\cos^2(\kappa z)\,
C_n^{(2)}\!\big(\sin(\kappa z)\big),
\end{equation}
where $C_n^{(2)}$ are Gegenbauer polynomials \cite{NIST:DLMF}. Beyond illustrating the general mechanism, the global $\mathrm{AdS}_2$ slicing for the value $\kappa = 1/(2\ell)$ provides a fully analytic fluctuation spectrum for a genuinely interacting field theory. This offers a rare example in which curvature, interactions, and stability can all be treated analytically within a nontrivial solitonic background.

It is important to emphasize that, within our BPS construction, the existence of the translational zero mode in global $\mathrm{AdS}_2$ is guaranteed by the general geometric criterion discussed above. This stands in contrast with the standard $\phi^4$ model formulated directly on $\mathrm{AdS}_2$ without the nonminimal curvature coupling, where no zero mode is present and the lowest excitation occurs at finite frequency, as shown in Ref.~\cite{VELDHUIS}. In our framework, the preservation of the flat-space BPS equation ensures that the zero mode survives whenever the foliation admits two asymptotic boundaries, despite the absence of translational isometries.

\begin{figure}[!h]
    \centering
    \includegraphics[width=0.48\columnwidth]{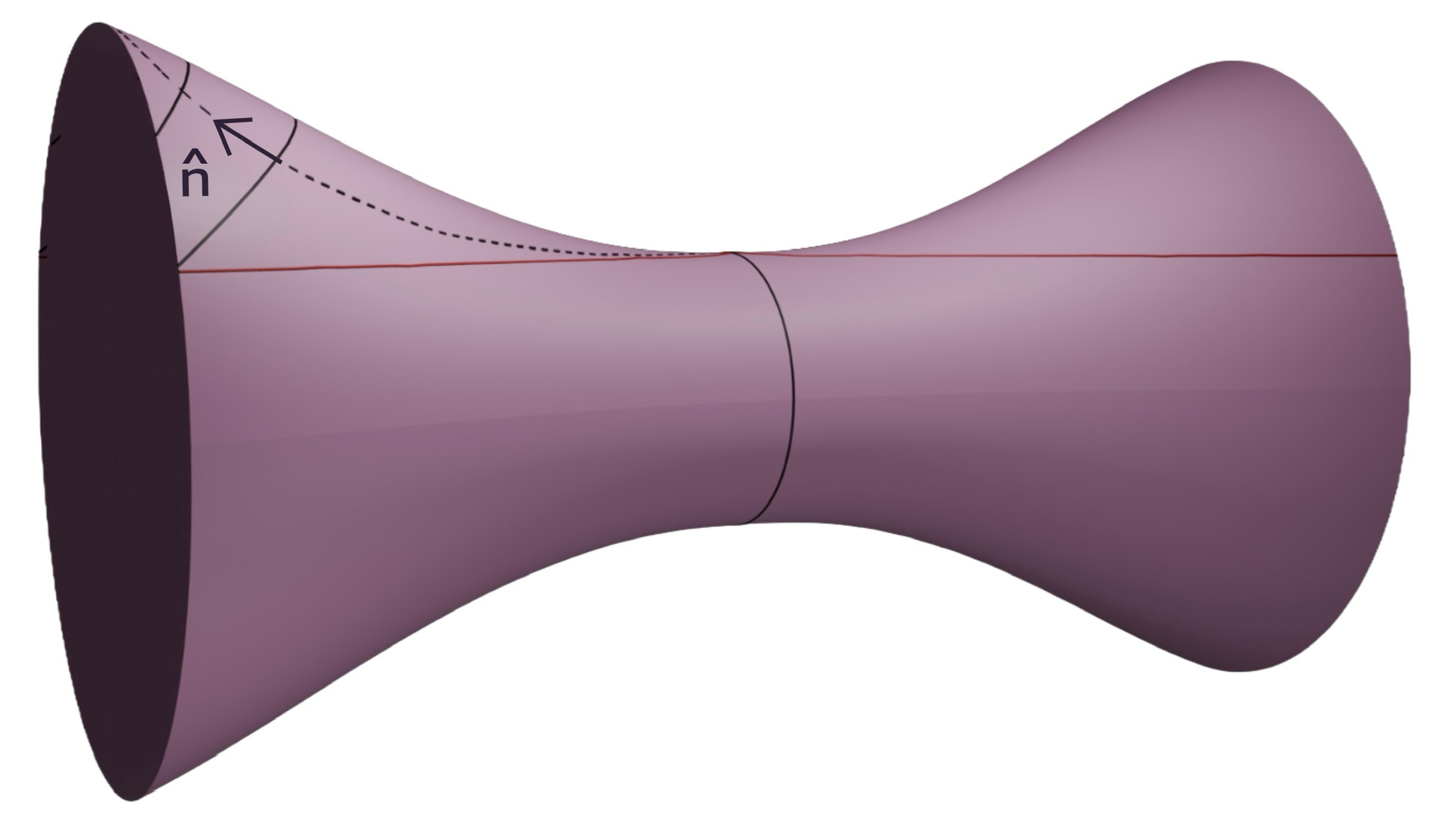}\hfill
    \includegraphics[width=0.48\columnwidth]{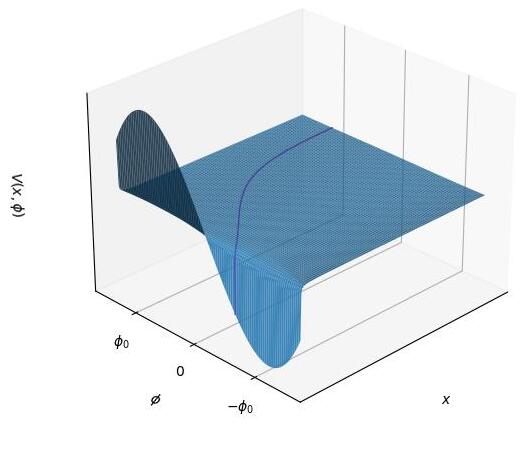}
    \caption{
Horizon-type slicing of the $\mathrm{AdS}_2$ hyperboloid with $h(x)=\sinh(\kappa x)$ and the corresponding BPS kink profile.
Although the kink remains an exact solution of the Bogomolny equation, the presence of a horizon truncates the spatial domain.
}

    \label{fig4}
\end{figure}

In the slicing defined by \eqref{case2}, the lapse function vanishes at $x\to0^+$ as $h(x)\sim \kappa x$, defining a
horizon-like boundary of the coordinate patch, which is depicted in Fig.\eqref{fig4}.

As discussed in the general analysis, the linear vanishing of the lapse function near
$x\to0^+$ implies that the translational zero mode is not square integrable in this slicing.
The presence of a horizon therefore removes the translational modulus independently of
any further details of the kink profile.

Instead, this slicing provides an instructive example in which the fluctuation spectrum can be analyzed exactly. At the tuned value $\kappa=1/(2\ell)$ the fluctuation potential is
\begin{equation}
\mathcal V(z)=-\frac{6\kappa^2}{\cosh^2(\kappa z)}
+\frac{2\kappa^2}{\sinh^2(\kappa z)}.
\end{equation}
This singular hyperbolic P\"oschl--Teller potential \cite{PT_original,Uly} is defined on the half-line and supports a purely continuous spectrum. The corresponding eigenfunctions can be written in terms of hypergeometric functions \cite{NIST:DLMF} as
\begin{eqnarray}
\psi_{\omega}(z)
= &&C_{\omega}\,
\big(\cosh(\kappa z)\big)^{-i\omega/\kappa}\times\nonumber\\
&&{}_2F_1\!\left(
i\frac{\omega}{2\kappa},
\frac{5}{2}+i\frac{\omega}{\kappa};
\frac{5}{2};
\tanh^2(\kappa z)
\right),
\end{eqnarray}
where $\omega^2$ labels the continuum of scattering states and $C_{\omega}$ is a normalization constant.

This explicitly solvable example highlights a central aspect of our construction: while the BPS kink remains linearly stable and the fluctuation operator preserves its supersymmetric factorization, the spectral content is sensitive to the choice of static foliation of $\mathrm{AdS}_2$. In the horizon-type slicing, the presence of a boundary at finite proper distance eliminates bound states, including the translational zero mode. This spectral rearrangement is therefore a global geometric effect tied to the chosen foliation, rather than a dynamical instability.

It is worth stressing that, although both the global and horizon slicings admit the tuned relation $\kappa\ell=1/2$ and lead to exactly solvable fluctuation problems, they define
inequivalent models. The corresponding effective potentials differ because the curvature coupling depends explicitly on the lapse function $h(x)$.  The two constructions are physically distinct, even though they are defined on the same underlying $\mathrm{AdS}_2$ spacetime.

Finally, we consider the static foliation defined by $h(x)=e^{\kappa x}$, for which the extrinsic curvature is constant. In terms of $z$-coordinate, the effective Schr\"odinger potential has the form
\begin{eqnarray}
\mathcal{V}(z)
=&&\frac{-\Delta}{z^2\big((\kappa z)^{\Delta}+1\big)^2}
\Bigg(-\Delta\big[(\kappa z)^{\Delta}-1\big]^2
+\nonumber\\
&&(\kappa z)^{2\Delta}-1
+2\Delta(\kappa z)^{\Delta}\Bigg), \ \Delta=\frac{1}{\kappa \ell}.
\end{eqnarray}

This slicing interpolates between an asymptotic $\mathrm{AdS}_2$ boundary and a horizon. Since the lapse function is exponential over the entire spatial domain, the normalizability of the translational zero mode is controlled by a single parameter $\Delta=1/(\kappa\ell)$. A direct analysis of the fluctuation measure shows that the zero mode is square integrable if and only if $\Delta>1/2$; for $\Delta\leq 1/2$ it is excluded from the physical Hilbert space. The existence of the zero mode is therefore determined by the competition between the intrinsic kink scale and the curvature scale of $\mathrm{AdS}_2$. Although no closed-form analytic solution of the fluctuation equation is available in this case, the existence or absence of the zero mode follows unambiguously from the general asymptotic analysis.

\begin{figure}[!h]
    \centering
    \includegraphics[width=0.48\columnwidth]{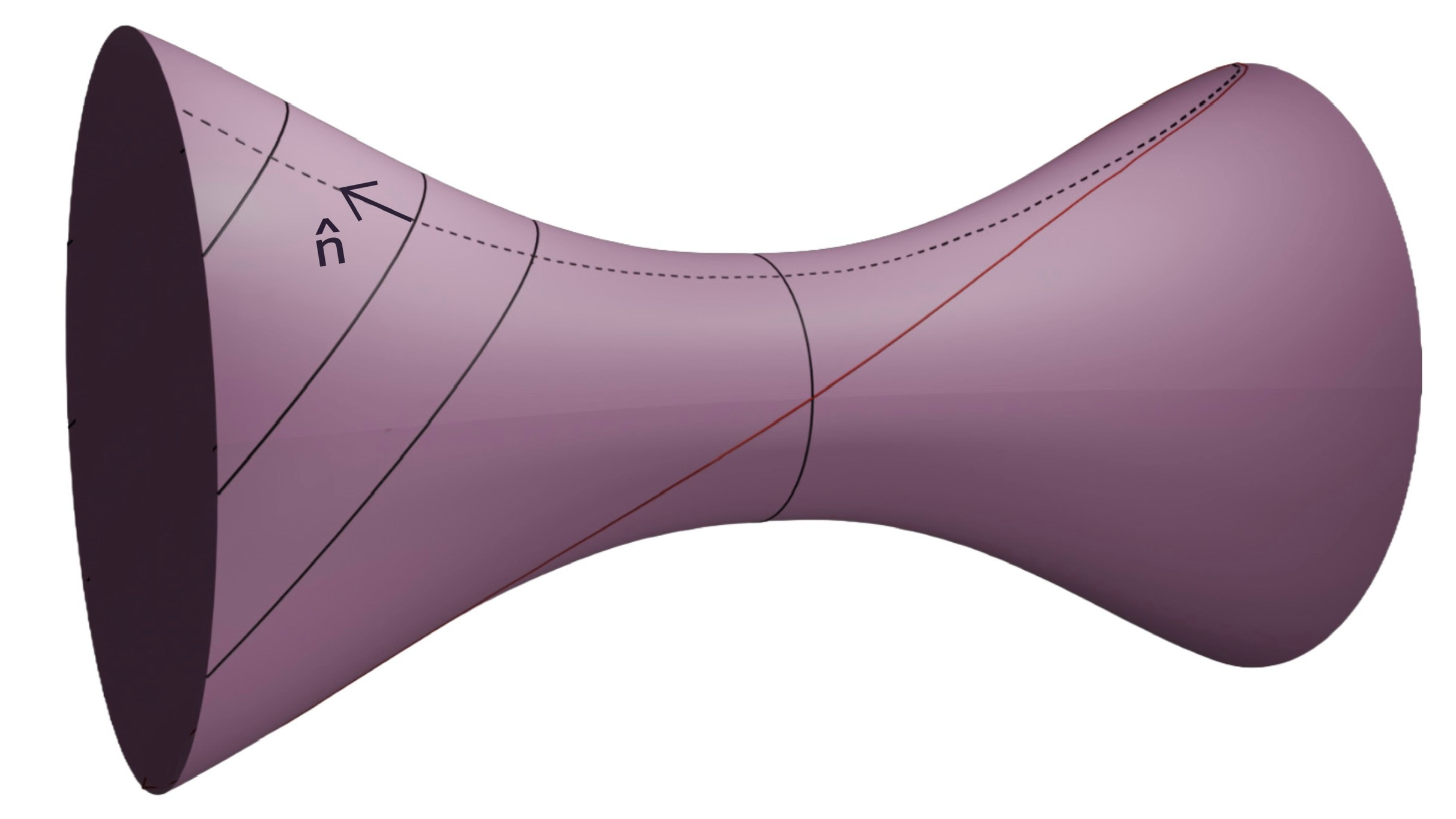}\hfill
    \includegraphics[width=0.48\columnwidth]{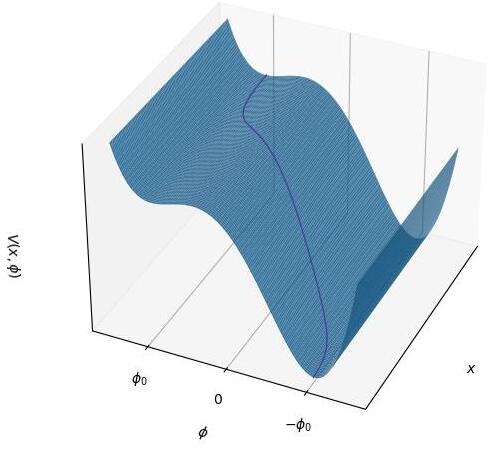}
    \caption{
Poincar\'e slicing of the $\mathrm{AdS}_2$ hyperboloid defined by $h(x)=e^{\kappa x}$ and the associated BPS kink profile. Geometrically, this slicing covers one half of the $\mathrm{AdS}_2$ hyperboloid, extending from an asymptotic boundary to a Killing horizon associated with the chosen static foliation.
}

    \label{fig5}
\end{figure}

Taken together, the three static slicings of $\mathrm{AdS}_2$ provide a unified geometric picture of how foliation and curvature control the fate of the translational mode. In the global slicing, Fig.~\ref{fig3}, the kink interpolates between two asymptotic boundaries and can be displaced without encountering geometric obstructions, ensuring the existence of a normalizable zero mode. In the horizon-type slicing, Fig.~\ref{fig4}, the coordinate patch truncates the soliton: any translation necessarily pushes part of the configuration beyond the horizon, and the zero mode is absent. The exponential slicing, Fig.~\ref{fig5}, represents an intermediate situation in which the kink extends along the hyperboloid but terminates at a single horizon. In this case, the survival of the zero mode is controlled by the competition between the intrinsic kink scale and the curvature scale of $\mathrm{AdS}_2$, illustrating how geometric effects alone can lift or preserve solitonic moduli without affecting linear stability.

\emph{Conclusions and outlook.—}
In this Letter we identified a universal geometric mechanism that extends the Bogomolny--Prasad--Sommerfield construction for scalar kinks to static curved spacetimes in $(1+1)$ dimensions. The key ingredient is a symmetry-preserving nonminimal coupling between the scalar prepotential and the extrinsic curvature of a static foliation. This coupling allows the \emph{flat-space} first-order Bogomolny equation to remain exactly valid in an arbitrary static background, independently of the intrinsic curvature of the spacetime. As a result, the kink profile is unchanged, while the potential is deformed in a controlled, purely geometric way determined by the lapse function $h(x)$ and the chosen static slicing.

Beyond the existence of the BPS configuration, we analyzed its small fluctuations. We showed that the spectrum is governed by a supersymmetric quantum-mechanical Hamiltonian, ensuring linear stability of the kink in any background compatible with our construction; in particular, for $\mathrm{AdS}_2$ this implies that the Breitenlohner--Freedman bound is automatically satisfied \cite{BF1,BF2}. Curvature nonetheless has a dynamical effect on the zero-energy sector: the translational mode that is always normalizable in Minkowski space may become non-normalizable in curved space. Its fate is controlled by a simple geometric criterion that compares the intrinsic kink scale (set by the vacuum mass) to the asymptotic geometric scale of the foliation (boundary versus horizon behavior of $h(x)$). When the criterion fails, the translational modulus is lifted and the kink becomes geometrically pinned, even though the BPS equation and the linear stability of the configuration remain intact. Explicit realizations in $\mathrm{AdS}_2$ illustrated how distinct static slicings of the same spacetime can lead to qualitatively different spectra - from preserved translational invariance to complete removal of the zero mode - highlighting that it is the interplay between curvature and foliation, rather than curvature alone, that governs the fate of solitonic moduli.

Our construction opens several natural directions. A first extension is essentially straightforward: the geometric mechanism is formulated in terms of the extrinsic curvature of a static foliation, and therefore admits a direct generalization to $(d+1)$ dimensions for backgrounds of domain-wall type, where the metric is static and warped along a distinguished spatial coordinate. In such settings, one expects the same logic to preserve flat-space first-order BPS equations.

A more ambitious direction is to extend the present construction to BPS solitons involving gauge fields. In particular, it is natural to ask whether known BPS  vortex solutions in Minkowski space~\cite{Bogomolny,JackiwWeinberg1990,HongKimPac1990} can be transferred, in a symmetry-preserving way, into curved spacetimes. Of special interest are asymptotically $\mathrm{AdS}$ geometries, where such configurations may have natural applications in holographic settings involving charged condensates~\cite{HartnollPRL,HartnollJHEP}. The central challenge is then to identify suitable curvature couplings—now involving both geometric and gauge-sector structures—that preserve the first-order BPS equations while ensuring a well-defined fluctuation problem.

A further natural application of our framework exploits the maximal symmetry of $\mathrm{AdS}$ spacetimes. Much like Minkowski space, $\mathrm{AdS}$ admits conformal boosts that map static solitonic configurations into genuinely time-dependent ones. As a consequence, once static BPS kinks are constructed in a given foliation, the underlying symmetry allows them to be promoted to dynamical solutions. This opens the possibility of constructing multi-kink configurations with large spatial separation and studying their boosted dynamics, including kink--kink and kink--antikink collisions \cite{CAMPBELL19831,Gabi_2}, directly in curved $\mathrm{AdS}$ backgrounds. Our construction therefore provides a controlled starting point for investigating nonperturbative soliton dynamics and scattering processes in maximally symmetric curved spacetimes.

\vspace{0.5cm}

\emph{Acknowledgments.—}
We are grateful to Karen Oliosi and Yuri Ribeiro de Carvalho for their assistance in the preparation of the figures.

\bibliographystyle{apsrev4-2}
\bibliography{refs_prl} 

@article{fake_super,
  author        = {D. Z. Freedman and C. Núñez and M. Schnabl and K. Skenderis},
  title         = {Fake Supergravity and Domain Wall Stability},
  journal       = {Phys. Rev. D},
  volume        = {69},
  pages         = {104027},
  year          = {2004},
  eprint        = {hep-th/0312055},
  archivePrefix = {arXiv},
  primaryClass  = {hep-th}
}

@article{PRL_Sken_Town,
  author        = {Kostas Skenderis and Paul K. Townsend},
  title         = {Hidden supersymmetry of domain walls and cosmologies},
  journal       = {Phys. Rev. Lett.},
  volume        = {96},
  pages         = {191301},
  year          = {2006},
  eprint        = {hep-th/0602260},
  archivePrefix = {arXiv},
  primaryClass  = {hep-th}
}

@misc{Freed_Gub,
      title={Renormalization Group Flows from Holography--Supersymmetry and a c-Theorem}, 
      author={D. Z. Freedman and S. S. Gubser and K. Pilch and N. P. Warner},
      year={1999},
      eprint={hep-th/9904017},
      archivePrefix={arXiv},
      primaryClass={hep-th},
      url={https://arxiv.org/abs/hep-th/9904017}, 
}

@book{Rajaraman,
  title={Solitons and Instantons; An Introduction to Solitons and Instantons in Quantum Field Theory},
  author={Rajaraman, R.},
  url={https://books.google.com.br/books?id=eiRqwgEACAAJ},
  year={1982},
  publisher={Amsterdam}
}

@book{MantonSutcliffe,
    author = "Manton, N.S. and Sutcliffe, P.",
    title = "{Topological solitons}",
    doi = "10.1017/CBO9780511617034",
    isbn = "978-0-521-04096-9, 978-0-521-83836-8, 978-0-511-20783-9",
    publisher = "Cambridge University Press",
    series = "Cambridge Monographs on Mathematical Physics",
    year = "2004"
}

@article{Cooper,
   title={Supersymmetry and quantum mechanics},
   volume={251},
   ISSN={0370-1573},
   url={http://dx.doi.org/10.1016/0370-1573(94)00080-M},
   DOI={10.1016/0370-1573(94)00080-m},
   number={5–6},
   journal={Physics Reports},
   publisher={Elsevier BV},
   author={Cooper, Fred and Khare, Avinash and Sukhatme, Uday},
   year={1995},
   month=jan, pages={267–385} }

@article{Bogomolny,
    author = "Bogomolny, E. B.",
    title = "{Stability of Classical Solutions}",
    reportNumber = "PRINT-76-0543 (LANDAU-INST.)",
    journal = "Sov. J. Nucl. Phys.",
    volume = "24",
    pages = "449",
    year = "1976"
}

@article{PrasadSommerfield,
  author  = {Prasad, M. K. and Sommerfield, C. M.},
  title   = {Exact Classical Solution for the ’t Hooft Monopole and the Julia-Zee Dyon},
  journal = {Phys. Rev. Lett.},
  volume  = {35},
  pages   = {760},
  year    = {1975}
}

@article{Agostinho,
  author  = {Adam, C. and Ferreira, L. A. and da Hora, E. and Wereszczynski, A. and Zakrzewski, W. J.},
  title   = {Some aspects of self-duality and generalised BPS theories},
  journal = {Journal of High Energy Physics},
  volume  = {2013},
  pages   = {62},
  year    = {2013},
  doi     = {10.1007/JHEP08(2013)062},
}

@article{Bazeia,
  title = {Deformed defects},
  author = {Bazeia, D. and Losano, L. and Malbouisson, J. M. C.},
  journal = {Phys. Rev. D},
  volume = {66},
  issue = {10},
  pages = {101701},
  numpages = {5},
  year = {2002},
  month = {Nov},
  publisher = {American Physical Society},
  doi = {10.1103/PhysRevD.66.101701},
  url = {https://link.aps.org/doi/10.1103/PhysRevD.66.101701}
}

@misc{NIST:DLMF,
         key = "{\relax DLMF}",
       title = "{\it NIST Digital Library of Mathematical Functions}",
howpublished = "\url{https://dlmf.nist.gov/}, Release 1.2.4 of 2025-03-15",
         url = "https://dlmf.nist.gov/",
        note = "F.~W.~J. Olver, A.~B. {Olde Daalhuis}, D.~W. Lozier, B.~I. Schneider,
                R.~F. Boisvert, C.~W. Clark, B.~R. Miller, B.~V. Saunders,
                H.~S. Cohl, and M.~A. McClain, eds."}

@article{Gabi_1,
  title = {Real scalar field kinks and antikinks and their perturbation spectra in a closed universe},
  author = {Hartmann, Betti and Luchini, Gabriel and Constantinidis, Clisthenis P. and Pereira, Carlos F. S.},
  journal = {Phys. Rev. D},
  volume = {101},
  issue = {7},
  pages = {076004},
  numpages = {11},
  year = {2020},
  month = {Apr},
  publisher = {American Physical Society},
  doi = {10.1103/PhysRevD.101.076004},
  url = {https://link.aps.org/doi/10.1103/PhysRevD.101.076004}
}

@article{Gabi_2,
  author = {Carlos F S Pereira et al},
  journal = {J. Phys. A: Math. Theor},
  year = {2021},
volume = {54},
  pages = {075701},
}

@article{Uly,
title = {Renormalization group and spectra of the generalized P\"osch-Teller potential},
journal = {Annals of Physics},
volume = {460},
pages = {169549},
year = {2024},
issn = {0003-4916},
doi = {https://doi.org/10.1016/j.aop.2023.169549},
url = {https://www.sciencedirect.com/science/article/pii/S0003491623003512},
author = {Ulysses {Camara da Silva} and Carlos F.S. Pereira and Andre {Alves Lima}},
}

@misc{witten1998antisitterspaceholography,
      title={Anti De Sitter Space And Holography}, 
      author={Edward Witten},
      year={1998},
      eprint={hep-th/9802150},
      archivePrefix={arXiv},
      primaryClass={hep-th},
      url={https://arxiv.org/abs/hep-th/9802150}, 
}

@article{Maldacena,
  author  = {Maldacena, Juan},
  title   = {The Large-N Limit of Superconformal Field Theories and Supergravity},
  journal = {International Journal of Theoretical Physics},
year    = {1999},  
volume  = {38},
number = {4},  
pages   = {1113--1133},
}

@book{WaldGR,
  author    = {Wald, Robert M.},
  title     = {General Relativity},
  publisher = {University of Chicago Press},
  address   = {Chicago},
  year      = {1984}
}

@article{Jackiw1985,
  author  = {Jackiw, R.},
  title   = {Lower Dimensional Gravity},
  journal = {Nucl. Phys. B},
  volume  = {252},
  pages   = {343--356},
  year    = {1985},
  doi     = {10.1016/0550-3213(85)90448-1}
}

@article{BardeenHorowitz,
  author  = {Bardeen, J. M. and Horowitz, G. T.},
  title   = {The Extreme Kerr Throat Geometry: A Vacuum Analog of AdS$_2 \times S^2$},
  journal = {Phys. Rev. D},
  volume  = {60},
  pages   = {104030},
  year    = {1999},
  doi     = {10.1103/PhysRevD.60.104030},
  eprint  = {hep-th/9905099}
}

@article{Krumhansl1975,
  author  = {Krumhansl, J. A. and Schrieffer, J. R.},
  title   = {Dynamics and Statistical Mechanics of a One-Dimensional Model Hamiltonian for Structural Phase Transitions},
  journal = {Phys. Rev. B},
  volume  = {11},
  pages   = {3535--3545},
  year    = {1975},
  doi     = {10.1103/PhysRevB.11.3535}
}

@article{BF1,
  author  = {Breitenlohner, Peter and Freedman, Daniel Z.},
  title   = {Stability in Gauged Extended Supergravity},
  journal = {Annals of Physics},
  volume  = {144},
  number  = {2},
  pages   = {249--281},
  year    = {1982},
  doi     = {10.1016/0003-4916(82)90116-6}
}

@article{BF2,
  author  = {Breitenlohner, Peter and Freedman, Daniel Z.},
  title   = {Positive Energy in Anti--de Sitter Backgrounds and Gauged Extended Supergravity},
  journal = {Physics Letters B},
  volume  = {115},
  number  = {3},
  pages   = {197--201},
  year    = {1982},
  doi     = {10.1016/0370-2693(82)90643-8}
}

@article{VELDHUIS,
author = {TER VELDHUIS, TONNIS},
title = {SOLITONS IN TWO-DIMENSIONAL ANTI-DE SITTER SPACE},
journal = {International Journal of Modern Physics A},
volume = {25},
number = {02n03},
pages = {289-299},
year = {2010},
doi = {10.1142/S0217751X10048615},

URL = { 
    
        https://doi.org/10.1142/S0217751X10048615
    
    

},
eprint = { 
    
        https://doi.org/10.1142/S0217751X10048615
    
    

}
}

@article{JackiwWeinberg1990,
  author  = {Jackiw, R. and Weinberg, E. J.},
  title   = {Self-Dual Chern--Simons Vortices},
  journal = {Physical Review Letters},
  volume  = {64},
  pages   = {2234--2237},
  year    = {1990},
  doi     = {10.1103/PhysRevLett.64.2234}
}

@article{HongKimPac1990,
  author  = {Hong, J. and Kim, Y. and Pac, P. Y.},
  title   = {Multivortex Solutions of the Abelian Chern--Simons--Higgs Theory},
  journal = {Physical Review Letters},
  volume  = {64},
  pages   = {2230--2233},
  year    = {1990},
  doi     = {10.1103/PhysRevLett.64.2230}
}

@article{AdS_2_holo,
  author = "Sarosi, Gabor",
  title = "{AdS$_{2}$ holography and the SYK model}",
  doi = "10.22323/1.323.0001",
  journal = "PoS",
  year = 2018,
  volume = "Modave2017",
  pages = "001"
}

@article{Uly_2,
author = {U Camara da Silva},
year = {2025},
journal = {J. Phys. A: Math. Theor.},
volume = {58},
number = {50},
doi = {10.1088/1751-8121/ae2446},
}

@article{PT_original,
author = {P{\"o}schl G and Teller E},
year = {1033},
title = {Bemerkungen zur quantenmechanik des anharmonischen oszillators},
journal = {Z. Phys.},
volume = {83},
pages = {143-51},
}

@article{CAMPBELL19831,
title = {Resonance structure in kink-antikink interactions in $\varphi^4$ theory},
journal = {Physica D: Nonlinear Phenomena},
volume = {9},
number = {1},
pages = {1-32},
year = {1983},
issn = {0167-2789},
doi = {https://doi.org/10.1016/0167-2789(83)90289-0},
url = {https://www.sciencedirect.com/science/article/pii/0167278983902890},
author = {David K. Campbell and Jonathan F. Schonfeld and Charles A. Wingate},
}

@article{HartnollPRL,
  title = {Building a Holographic Superconductor},
  author = {Hartnoll, Sean A. and Herzog, Christopher P. and Horowitz, Gary T.},
  journal = {Phys. Rev. Lett.},
  volume = {101},
  issue = {3},
  pages = {031601},
  numpages = {4},
  year = {2008},
  month = {Jul},
  publisher = {American Physical Society},
  doi = {10.1103/PhysRevLett.101.031601},
  url = {https://link.aps.org/doi/10.1103/PhysRevLett.101.031601}
}

@article{HartnollJHEP,
  author       = {Hartnoll, Sean A. and Herzog, Christopher P. and Horowitz, Gary T.},
  title        = {Holographic Superconductors},
  journal      = {JHEP},
  volume       = {12},
  pages        = {015},
  year         = {2008},
  doi          = {10.1088/1126-6708/2008/12/015},
}
\end{document}